# Energy Efficient LoRaWAN in LEO Satellites

Muskan Shergill
College of Engineering
*The Ohio State University*
Columbus, OH, USA
shergill.10@buckeyemail.osu.edu

Zach Thompson
College of Engineering
*The Ohio State University*
Columbus, OH, USA
thompson.4012@buckeyemail.osu.edu

Guanqun Song
College of Engineering
*The Ohio State University*
Columbus, OH, USA
song.2107@osu.edu

Ting Zhu
College of Engineering
*The Ohio State University*
Columbus, OH, USA
zhu.3445@osu.edu

*Abstract* — LPWAN services' inexpensive cost and long-range capabilities make it a promising addition and countless satellite companies have started taking advantage of this technology to connect IoT users across the globe. However, LEO satellites have the unique challenge of using rechargeable batteries and green solar energy to power their components. LPWAN technology is not optimized to maximize battery lifespan of network nodes. By incorporating a MAC protocol that maximizes node the battery lifespan across the network, we can reduce battery waste and usage of scarce Earth resources to develop satellite batteries.

*Keywords—LEO Satellites, LEO, LoRa, LPWAN, Energy, Efficiency, Energy-Efficient, MAC Protocol*

## I. Introduction

In this paper, we build upon an existing energy-efficient LoRa protocol designed for terrestrial IoT devices to handle the unique constraints of LEO satellites. We adopted a popular and accurate degradation model [9] [10] for lithium-ion batteries and generalized it to meet the dynamic constraints of LEO satellite batteries. We modified the battery-aware MAC protocol to consider the unique green energy usage in space and satellite-focused forecast windows. The simulation used was one borrowed from a classmate. It simulates the travel of satellites and the connections they form, maintain and lose based on proximity to each other and to ground stations. While not immediately useful, any future work could very much benefit from using the simulation as a predictive basis for LoRa.

## II. Background

The growth of the Internet of Things (IoT) has driven a surge in demand for low-power, wide-area networks (LPWANs). These wireless communication technologies are designed to connect battery-operated devices to the Internet over long distances at regional, national, and even global scales [1]. By utilizing lightweight protocols and cost-effective components, LPWANs minimize complexity and reduce deployment costs. Their key features of low power consumption and long-range capabilities make them ideal for diverse applications such as smart agriculture, healthcare, and urban infrastructure management [3][5]. In addition, they are incredibly scalable, allowing hundreds of devices to connect to their gateways, enabling grand connectivity [5]. LoRa (Long Range), a key LPWAN specification, utilizes modulation techniques at the physical layer to encode data onto wireless carrier waves. LoRaWAN leverages this modulation to manage communications between LoRa gateways and end-node devices [5]. Applications that use LoRa do not use a continuous flow of data communication, giving each device a "turn to speak" and prevents congestion at the gateway [5]. Together, these components make up a complete LoRaWAN solution.

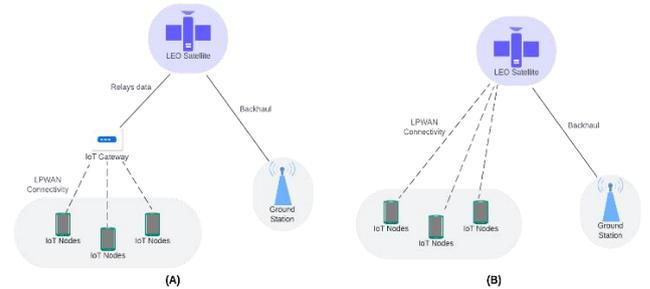

Fig. 1: IoT – Satellite Communicaiton Architecture
(A) ItS-IoT and (B) DiS-IoT

Looking beyond terrestrial infrastructure, LoRaWAN and other LPWAN technologies are being used in space-based IoT systems. IoT satellites have the proper aspects to support LPWAN architecture. They facilitate connectivity among IoT devices and transmit small data to terrestrial infrastructure or other satellites via sensors (or nodes) [8]. By design, they are oriented toward wide-area networks to support coverage to various geographical areas and are scalable to support large number of devices [8]. Following these advantages, these satellites can follow two communication architectures as illustrated in Figure 1. Direct-to-Satellite (DiS-IoT) allows IoT nodes to communicate directly with Low Earth Orbit (LEO) satellites [8]. Alternatively, Indirect-to-Satellite (ItS-IoT) employs intermediate gateways to relay data to satellites using standard protocols like those defined by the Consultative Committee for Space Data Systems (CCSDS) [8]. Companies like Lacuna Space, Thuraya, and Wyld Networks use LoRa connectivity in their developments while SatelIot, Ligado, and GateHouse use NB-IoT [8].

However, LoRaWAN in satellites does not come without challenges. LPWANs are not equipped to maximize the battery lifespan of the network nodes according to [10]. LoRa [1] depends on ALOHA MAC protocol which sends packets immediately they are generated. While this can impact the longevity of IoT devices using rechargeable batteries or green energy, the issue is expedited in satellites. Low Earth Orbit (LEO) satellites rely on solar energy for their operations, processing, and communication when exposed to sunlight. Excess solar energy is stored in onboard batteries for use during periods of eclipse, where sunlight is unavailable. However, these batteries have finite charge-discharge cycles, known as the depth of discharge (DOD), which directly impacts their longevity. Prolonged stress on these batteries not only reduces their effective lifespan but also limits the operational life of the satellites the power [2]. We need to consider how to optimize these battery lifespans with LPWAN technologies to get the most out of LEO satellites.

In this paper, we make the following contributions:

- We adjusted an optimized lithium-ion battery degradation model to consider space environment and satellite components. These are generalized equations but aim to better demonstrate battery degradation unique to LEO satellites.
- We adjusted an On Sensor Forecast Algorithm to consider forecast windows for satellite visibility and energy availability based on resource constraints.
- We analyze a satellite travel simulation as it relates to a network between themselves and ground stations and consider how it could be used to aid LoRa in creating effective predictions on its nodes.

## III. RELATED WORKS

While LoRaWAN and other LPWAN technologies are primarily optimized for terrestrial applications, there is an increase of use in space-based systems such as LEO IoT satellites. Some of the leading companies offering LPWAN services include SpaceX (Starlink), Telesat, OneWeb, Airbus, and Lacuna Space [8]. This section discusses existing research that has influenced this study. Most satellite research on the topic involves LPWAN performance and data integrity in space networks.

The *Battery Lifespan-Aware Mac Protocol* defined in [10] proposed a localized decision-making approach to optimize battery usage for IoT nodes. It integrated battery degradation model for lithium-ion batteries and employed predictive energy management protocol based on SoC and DoD to improve battery lifespan in LoRaWAN systems. While this work was developed for terrestrial IoT networks, its principles aligned with the challenges of LEO satellites where energy management is critical. This paper adopts these principles and serves as a foundation for adapting the protocol to the unique requirements of LEO satellites.

The research in [11] provides detailed experimental results on the effects of thermal cycling and DoD battery longevity. We discuss these results in more depth in Section IV and apply these parameters to our experiment design.

Adrian Petrariu and fellow researchers developed a hybrid power management system [15] for LoRa communication using renewable energy. Their system integrated a photovoltaic (PV) panel, supercapacitor (SC), and a small lithium battery to improve energy efficiency, ensuring long-term functionality of IoT sensor nodes without compromising data transmission intervals. This work provides insight on when power is most consumed in LoRaWAN networks and how a green energy system can counteract these constraints.

## IV. SATELLITE DESIGN

### A. Satellite Thermal Control

One of the main challenges of space technology is maintaining operable internal temperature for internal components. The device's position from the sun, thermal management system, and indirect solar heating all contribute to the extreme space temperatures satellites can experience [13][14]. The following equation and Figure 3 are a simplified overview of how these factors are used to balance (heat) energy on spacecrafts [13].

$$q_{solar} + q_{albedo} + q_{shine} + Q_{gen} - Q_{out,rad} = Q_{stored} \quad (1)$$

$Q_{gen}$ = heat generated by the spacecraft
$q_{solar}$ = solar heating
$q_{albedo}$ = solar heating reflected by the planet
$q_{shine}$ = infared heating from the planet
$Q_{out,rad}$ = heat emitted via radiation
$Q_{stored}$ = heat stored by the spacecraft

These heat fluxes are dynamic due to satellite orbits. For LEOs, the heat stored should support an internal temperature range of -65°C to +125°C depending on their orbital height [14]. Lithium-ion batteries have a smaller range of operable temperatures. The battery specification in part *B. Battery Considerations* has an operating temperature range of -20°C to +40°C [11].

These temperatures can be regulated using passive, active, or both passive and active thermal management technology. Passive systems are good choices for smaller satellites like LEOs and SmallSats due to their low cost, volume, and weight. Passive technology options include [13]:

- *Sprayable Thermal Control Coatings*: liquid coating applied to surfaces to manage absorption and emission of energy
- *Thermal Straps*: conductive link between heat sources and thermal sink
- *Sunshields:* reduces amount of solar heating by blocking view to the sun
- *Thermal Switches:* controls current flow in response to temperature changes
- *Radiators*: dissipates excess heat via radiative heat transfer
- *Films, Tapes, and MLI:* surface materials to regulate temperatures

Using a combination of these various passive technologies vastly provides a lifeline for the batteries to maintain their temperatures. It's akin to the physical properties of a house, protecting it from the harsh outside simply through walls, doors, and insulation. With passive technologies we get our main insulation in the form of thermal coatings, tapes and MLI. To understand why this work we first need to understand that in a vacuum heat is only transferred by radiation and conduction. Conduction is usually not an issue in a fully enclosed small satellite, so that brings us to radiation, the main force behind heat transfer to and from the outside environment [13].

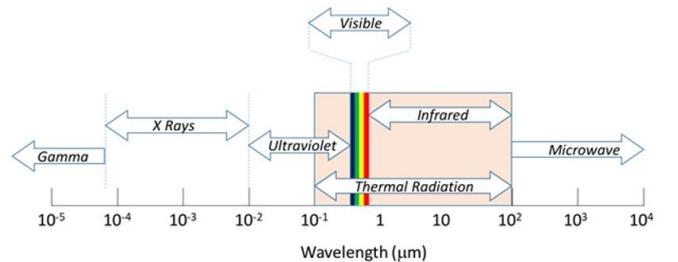

Fig. 2: Electromagnetic spectrum featuring the range of Thermal Radiation

Given that radiation is what mainly affects the internal temperatures of batteries, it stands to reason satellites try to employ coating that has specific optical properties to maintain temperatures: namely solar absorptivity (the amount of heating from solar radiation that is absorbed) and IR emissivity (how much heat is emitted out and into the environment). Generally, we want less IR emissivity and more solar absorptivity. Because of this we can use a specific sprayable coating that absorbs a lot of solar radiation, while also acting as an insulator, curbing IR emissivity. There may be some coatings that work too well though and can keep the internal temperature too hot. That said, there are many coatings to choose from such as different mixes of polyurethane, silicate, and silicones, and after testing the correct coating can be chosen.

Beyond thermal coatings, MLI blankets are incredibly instrumental in preserving heat. They are made up of 10-20 layers of material with low IR emissivity and an outer layer to protect the rest. However, MLI blankets are delicate and frankly not very effective in smaller form factors given the loss of efficiency at the edges of the blankets. They should be used with medium to large satellites. While they are not always used at the same time as sprayable thermal coatings, they certainly can be [13].

While retaining is incredibly important in space, internal workings on a satellite can get incredibly cold when not exposed to sunlight. For this we have sunshields and radiators among other things. Sunshields are essentially just a shade that blocks view to the sun made up of a material with low solar absorptivity. Radiators are just like they are in a car; they're material that is deployable or stowable to be used when there is excess heat that needs to go somewhere [13]. Using all these passive technologies we can help to both dissipate and retain heat for our batteries.

In contrast to passive technology, active technology requires a power source to work but can better maintain tight temperature control to its passive counterpart. The following are active thermal technology options [13]:

- *Electrical Heaters*: polyamide film that produces heat when current is applied
- *Cryocoolers:* refrigeration system capable of cooling components to -173°C or below
- *Thermal Coolers (TEC):* heat pumps that use Peltier effect to provide localized cooling
- *Pumped Fluid Loops (PFL):* moves liquids through tubes to cool multiple locations

Both active heating and cooling are of incredible importance when it comes to batteries in LEO satellites, which is our main concern as it pertains to temperature regulation. Given that batteries are typically the component in satellites that are most sensitive to temperature, needing to stay in a narrow temperature range, not only would a good heater be necessary, but the intense heat from direct sunlight would make effective cooling systems just as important. Whether it be space heaters, refrigeration, heat pumps or liquid cooling systems, any and all can and should be used to preserve batteries and other components.

TABLE I: Li-Ion Battery Experimental Values

| Parameter | Value |
|---|---|
| T | 263K (eclipse) and 303K (sunlight) |
| R | 8.314 J/mol * K |
| DoD | 40% |
| c | 1.3 |
| N | 15-16 charge-discharge per day for 90-minute orbit. 5,840 cycles a year |
| $E_a$ | 35,000 – 40,000 J/mol |
| C-rate | 12.5 A |
| d | 1.2 |
| SoC | 75%-90% |
| b | 1.3 |
| $k_1$ | 5.5 X $10^{-3}$ |
| $k_2$ | 2.0 |

*B. Battery Considerations*

Batteries must have very high energy density, could withstand extreme temperature fluctuations, have long life spans, and be durable enough to survive extreme environmental factors such as vibrations, collisions, and radiation [6]. For decades, nickel-cadmium batteries (Ni-Cd) were a popular choice for LEO and GEO satellites since their lightweight and inexpensive. However, they're not as energy-dense and their high discharge rate can cause overheating [6][7]. Nickel-hydrogen (NiH$_2$) was developed as hybrid between fuel cells and battery technology to increase energy density and capacity. It improves the overheating issue, but it has low volumetric energy density and requires high pressure storage to hold the hydrogen gas generated during charging [6][7]. In recent times, lithium-ion (Li-Ion) batteries are the most used technology as they provide higher energy levels and longer cycle life at a lower weight than Ni-Cd or NiH$_2$ batteries [6]. They also can operate at lower temperature environments than their counterparts, essential for space applications [12]. Considering most LEO satellites use Li-Ion batteries, we will use those parameters and assumptions with our model and simulation.

Specifically, we refer to the publicly available publication [11] that evaluated the performance of two 28 V, 25 Ah lithium-ion batteries under LEO satellite missions. Each battery contained eight prismatic lithium-ion cells connected in series. LEO satellites are limited in mass and space, so the prismatic design allows for efficient packing and high energy density per unit volume [12]. Each cell had a mescocarbon microbeads (MCMB) anode, a lithium nickel cobalt oxide (LiNiC$_0$O$_2$) cathode, and a liquid organic electrolyte [11]. These design choices further ensure high energy density, thermal stability, and long cycle life. Table 1 outlines the key experimental values from [11] to be used in our experiment. The parameters correspond with the battery degradation equations in *C. Battery Degradation*.

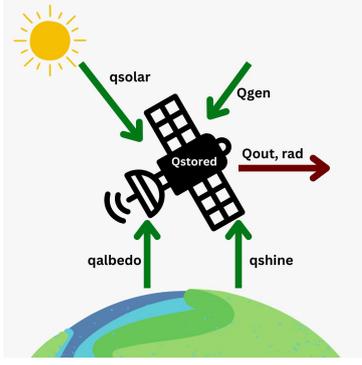

Fig. 3: Simplifed Overview of Heat Balancing for Satellite

## C. Battery Degradation Model

The model developed in [9] is considered highly accurate Li-Ion battery degradation model. It was also adopted in a related paper observing battery lifespan-aware protocols in terrestrial systems [10]. We took these models and provided general equations considering space environment and LEO satellite considerations.

First, calendar aging demonstrates the battery's inherent degradation over time based on average internal temperature and battery state of charge (SoC). Equation 2 is a generalized version of the calendar aging model from [9] and [10], using known parameters and experimental results from [11]. We incorporate the Arrhenius equation, which calculates the exponential factor that influences the rate of a chemical reaction based on the activation energy, ideal gas constant, and absolute temperature (for our case the average internal temperature) [16].

$$\Delta C_{cal} = k_1 * \exp\left(\frac{-E_a}{RT}\right) * (SoC)^b * t \quad (2)$$

$k_1$ = calibration constant
exp = exponential function
$E_a$ = activation energy of degradation
R = universal gas constant
T = average internal temperature
SoC = state of charge (0-1 scale)
b = exponent based on battery chemistry
t = time (in days)

Second, cycle aging demonstrates the life lost between charge and discharge cycles. It depends on the number of charge-discharge cycles [10][11], but we adapted it to also include the Arrhenius equation [16].

$$\Delta C_{cycle} = k_2 * (DoD)^d * (C)^c * \exp\left(\frac{-E_a}{RT}\right) * N \quad (3)$$

$k_2$ = calibration constant
C-rate = current rate
c,d = exponents specific for DoD and C
N = number of cycles

Combining Equation 2 and 3, the linear degradation for batteries is represented by the following equation:

$$\Delta C_{total} = \Delta C_{cal} + \Delta C_{cycle} \quad (4)$$

Lastly, we have the formula for Nonlinear SEI (Solid Electrolyte Interphase) formation. SEI film is something that forms on the surface of electrode material in lithium-ion batteries overtime. As it forms and it introduces a nonlinearity to the battery that causes a slight but irreversible loss of charge and output capacity. The final formula essentially takes the linear degradation formula and adds (or rather subtracts from 1) a component to factor in how SEI film formation effects degradation overtime in addition to linear degradation.

$$1 - \alpha_{sei} * \exp(-k * D_L) - (1 - \alpha_{sei}) * \exp(-D_L) \quad (5)$$

k = film formation constant
$D_L$ = linear degradation
$\alpha_{sei}$ = capacity lost due to SEI formation
exp = exponential function

With this final formula we can compute battery degradation overtime in any given system, and with the small tweaks more specifically in a satellite system.

## D. LoRaWAN

LoRaWAN networks take advantage of the star-of-star topology because it provides centralized control, scalability, and flexibility. The network architecture consists of the following components, also illustrated in Figure 4 [17][20]:

- *End Devices*: battery-operated sensors, IoT devices, that send or receive messages using LoRa RF modulation
- *Gateways*: forward messages to the network server
- *Network Server:* manages the entire LoRaWan network
- *Application Servers*: securely processes application data
- *Join Server*: processes join-request messages sent by end devices

LoRa RF modulation uses a technique called Chirp Spread Spectrum (CSS). It encodes data using chirp signals, and the frequency of the chirp can either increase or decrease over time [19][20]. The actual encoded data elements are "chips", and the number of chips per bit is called the spread factor or SF [19]. The larger the SF, the slower the data rate and vice versa. This may seem inefficient, but the slower data rate allows for longer potential communication range [17][20]. In the case of LEO satellites, higher SF values like SF10-SF12 would enable longer communication periods between the satellite and ground stations which is necessary considering the long distances between them. LoRa also employees Forward Error Correction (FEC) which adds redundant bits to improve receiver's sensitivity thus increasing reliability [19].

There are two types of LoRaWAN gateways identified by [17], picocell (indoor) and macrocell (outdoor) gateways. Picocells are ideal for locations with multiple obstacles or walls (i.e. basements, multi-floor buildings, etc.) while macrocells provides coverage to outdoor rural and urban areas [17]. We will assume macrocell gateways for wide coverage. LEO satellites can also act as a gateway.

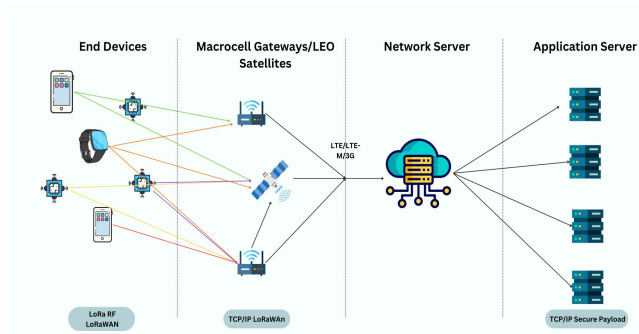

Fig. 4: LoRaWAN Network Architecture

The network server holds multiple responsibilities to manage all components in the LoRaWan network. This includes selecting the best gateway for downlink routing, address checking, acknowledgements for uplink messages, responding to MAC layer commands, and establishing 128-bit Advanced Encryption Standard (AES) connections for end-to-end security [17].

The join server receives the join-request message from the end device and processes the request by generating NwkSKey and AppSKey and transferring them to the network and application server respectively [17].

*E. Problem Formulation*

The study [10] provides a problem formulation that aims to maximize the battery lifespan of terrestrial devices in the LPWAN network. We adjust this formulation to ensure efficient operations of LoRaWAN in LEO satellites with a focus on (1) optimized battery lifespan, (2) balancing transmission efficiency with the challenges posed by LEO satellite dynamics such as Doppler shift, and (3) minimized energy wastage while maintaining high data utility.

The generalized battery degradation model in *Section IV-C* was already adjusted from the model in [10], but it should be noted that it doesn't account for the full range of operable battery temperatures, just the average.

Renewable energy sources are a key aspect of LEO functionality. Assuming the satellite takes 90 minutes to orbit the Earth, the spacecraft is exposed to the sun for 55 minutes and is eclipsed for the remaining 35 minutes [11]. The battery is charged for those 55 minutes and then discharges 40% of the capacity (10 Ah) during the 35-minute eclipse at a nominal discharge rate of 0.7C [11]. Therefore, forecast windows are optimized during sun exposure due to green energy generated, but then we have the challenge of 35-minutes of darkness. From a quality-of-service standpoint, the network must be able to transmit data during the eclipse period. This is where we dynamically check energy storage and battery state of charge to pick the most sustainable forecast window.

Like the terrestrial research [10], we define the transmission time slot for each node (satellite), the percentage of solar energy used for packet transmission, and battery recharging within each time slot. Since the battery is only charged during the eclipse period, not every time slot will have battery recharged value. Thus, for node u, we have two decision variables like in [10]:

- $x_u[t]$ $X_u$ is 1 if packet transmission happens during time slot (t), 0 otherwise
- $y_u[t]$ $Y_u$ is the value of solar energy used during time slot (t), either 0 or 1

Using these decision variables, we can express the battery degradation as $D_u(T, X_u, Y_u)$ where T are the time slots [10]. Then, we determine the energy generated $E_g$ at each node during time slot t as [10]:

$$\varphi_u[t] = \varphi_u[t-1] + y_u E_g[t] - x_u[t] E_{cons} - (1 - x_u[t]) E_{sleep}$$

(6)

$E_{cons}$ is the energy consumed during packet transmission while $E_{sleep}$ is energy consumed while node is asleep [10]. For simplicity, these equations follow the same problem formulation as [10]. However, since green energy generation is predictable and cyclic in satellites, we can safely say that $E_g$ is nonzero during sun exposure and 0 during eclipse phases.

For energy consumption of packet transmission in LoRa and the number of symbols in a packet, we use the Semtech LoRa Calculator [21] assuming the radio transceiver used is SX1262 due to its diverse application range and support of global frequency bands [22]. Alternatively, the energy consumption for packet transmission and symbols in a packet equation in [10] can be applied.

Overall, the problem formulation closely follows the terrestrial version in [10], making it incredibly easy to adapt the proposed MAC protocol to be compliant with LEO satellites.

*F. On Sensor Approach*

We use the foundation of the On-Sensor Approach for MAC protocol from [10]. We assume that each (satellite) node locally decides transmission time and solar energy usage and then uses a slotted-ALOHA approach by defining forecast windows for transmission. We assume that the nodes are not synchronized for simplicity, but this assumption may cause packet loss or collision issues in real-life implementation. To solve the battery lifespan maximization problem, we adapt the main challenges from [10] to satellites:

```
Algorithm 1: On-Sensor Forecast Window Selection
Inputs:
    – T: Set of all forecast windows t
    – T_sun: Set of forecast windows in sunlight phase
    – T_eclipse: Set of forecast windows in eclipse phase
    – Ψ: Current battery state of charge (SoC)
    – Ψ_max: max battery capacity
    – Ψ_min: min energy required
    – E_g: solar energy generated
    – E(t): Energy required for transmission at window t
    – E_critical: Energy needed during eclipse phase
Outputs:
    – Selected forecast window (t) for transmission OR packet drop
1. Initialize decision variable X_u to 0
2. For each forecast window t in T
    If t is in sunlit phase, estimate energy available
      If estimated energy ≥ Ψ_min + E_critical
        Compute DIF
        Compute weighted objective
        If objective is optimal for t
          Update Ψ
          RETURN T
      Else prioritize charging DROP PACKET
    If t is in the eclipse phase, estimate battery energy
      If Ψ_min > Ψ
        DROP PACKET
      Else
        RETURN T
3. Default
    DROP PACKET
```

(1) Estimation of green (solar) energy generated
(2) Packet collisions from unsynchronized nodes
(3) Sharing battery information amongst nodes

We define forecast windows as when the satellite is visible to ground devices (end devices or stations) or other satellites. These windows are the times when communication is possible between nodes. Additionally, the pattern of solar exposure is predictable and cyclic based on the satellite's orbital parameters. So, we do not have to consider random changes to solar energy and can focus on maximizing energy storage during sunlight phases and optimize battery usage during eclipses. We assume that satellite forecast windows range from a few minutes to maximum of 30 minutes, depending on the satellite's altitude and speed relative to the Earth [23]. Considering this time limit, we use the transmission parameters defined in [10]: 8 retransmissions of 10-byte packets take 40 seconds to finish with a spreading factor of 10. We could also apply a SF of 12 considering the distance challenges of satellites.

To compensate for packet collision, we use the Exponentially Weighted Moving Average (EWMA) equation [10]:

$$e_{ewma}[t] = \beta \cdot E_{cons}[t-1] + (1-\beta) \cdot e_{ewma}[t-1] \quad (7)$$

The goal is to give newer data precedence over older data. $e_{ewma}$ is the transmission energy estimate at t and $E_{cons}$ is the actual value from the previous time t − 1. $\beta$ is the precedence weight predetermined by the network manager.

Lastly, nodes have limited computing power and energy so calculating battery degradation at each node is too complex. Instead, nodes will send a summary of their battery usage to the gateway, and the gateway uses the information to calculate the degradation for all nodes. The methods used to handle this challenge are defined in [10].

Algorithm 1 is the adopted pseudo-code to selecting forecast window at a given node (satellite). It takes input $T_{sun}$ and $T_{eclipse}$ (set of forecast windows in sunlight and eclipse phase). As stated previously, this set is predictable and cyclic. Ψ, $Ψ_{max}$, and $Ψ_{min}$ define the battery's capacity parameters. Lastly, various energy variables to track available energy at various phases. We first initialize the decision variable $X_u$. Depending on if the forecast window is in sunlight or eclipse phase, we estimate either green energy generated or available battery level. The *Degradation Impact Factor* or DIF aims to estimate the impact of transmitting in window t on degradation. It follows Equation 15 in [10] and the result is a number between [0,1]. A DIF[t] > 0 means degradation increases due to cycle aging and DIF[t] = 0 means there is negligible impact on cycle aging. The objective function in [10] aims to minimize energy consumption and maximize packet transmission. If it's optimal to transmit the packet in the respective phase, the node transmits in the particular forecast window t. Otherwise, the packet is dropped, and we wait for a more optimal forecast window.

*G. Simulation*

The simulation we mainly used was not of our own design but rather borrowed from our classmate Keegan Sanchez. While it does not explicitly provide the information we want, it is instrumental in showing off the possibilities of LoRa and how smooth the technology can be in keeping satellites connected at all times.

The simulation itself is essentially an animated display of Starlink satellites and their connections with each other, as well as their connections with various ground stations. As the satellites move you can see where the connections are maintained, where they are dropped, and where new connections are formed. For the most part connections between satellites are maintained, with many moving in parallel with those they relate to. The differences lie with the ground stations, which are constantly dropping and adding connections as satellites are coming in and out of them.

In Figure 5 you can see red nodes, all representing satellites, and green nodes representing ground stations. From those you have the red lines, which are connections between two satellites and green lines which are connections between a satellite and a ground station. As the satellites move around the earth, they form new connections with stations and drop connections as they get too far away. There were some issues with running the simulation as it tended to freeze after 30 seconds when it is run, however it provides a good for which to observe the behavior of these satellite networks.

We would have liked to edit the source code to more directly address the LPWAN technologies that would be used here. This includes battery degradation parameters for lithium-ion batteries, solar energy generated based on satellite orbits, LoRaWAN parameters (i.e. SF, bandwidth, etc.) That said, there were continuous issues when trying to edit Keegan's simulation beyond what he had originally planned for it. Regardless, while not immediately useful to our purpose, the simulation provides an incredibly valuable model that could be used in the future as a predictive model for the LPWAN devices. The code for this simulation can be found at [18].

## V. IMPROVEMENTS/FUTURE WORK

There are many improvements that could be made for the future of this idea. As for future work, another simulation modeling battery degradation given all dynamic variables such as, environment, sunlight, solar panel model, battery model, etc. would be a great next step. Once finished, merging the battery simulation with the satellite travel simulation can provide accurate values regarding green energy generated, finetuned sun vs eclipse windows based on unique orbits, and packet transmission from satellite to satellite or satellite to ground station. Being able to track battery degradation as it relates to connections being made between satellites and ground stations would provide direct insight into how the connections made affect battery degradation and how we could possibly remedy that issue, or if it even is an issue that should be addressed at all.

If we determine that battery degradation is a concern for LEO satellites using LoRaWAN technology, one potential solution would be to implement adaptive transmission protocols. These protocols could adjust transmission intervals and spreading factors dynamically, depending on satellite visibility, orbit height, and contact duration, to maximize battery lifespan.

Alternatively, if the impact of transmission protocols on battery life is minimal, we can focus on further optimization of LoRaWAN services in satellite communications. For instance, the Doppler effect plays a significant role in communication constraints in satellite systems. Developing a Doppler-tolerant modulation scheme could help reduce transmission errors caused by the relative motion of satellites.

In the future, we can also incorporate nowadvanced research directions, such as wireless networks [24-41], secure communications [42-45], and machine learning [46-55]. We can use machine learning to optimize battery management and satellite communications, for example by tracking battery degradation in real time through predictive models, and things like dynamically adjusting transmission protocol parameters to balance communication needs and energy consumption. In addition, the application of secure communication technologies and advances in wireless networks (e.g., LoRaWAN mesh networks) can further improve the reliability and efficiency of satellite systems.

## VI. CONCLUSION

While the current simulation does not encapsulate our full design, it provides a valuable starting point for understanding the behavior of satellite networks. Future work should focus on creating the forecast window sets for sunlight and eclipse phases, applying battery lifespan maximization, and incorporate the on-sensor approach. With these improvements and analysis, we hope to improve battery and resource sustainability.

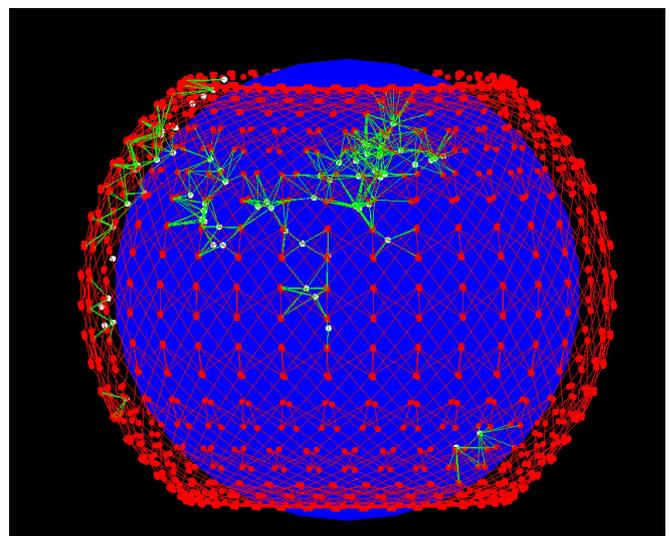

FIG. 5: SATELLITE CONNECTION SIMULATION